\documentclass[twocolumn,preprintnumbers,superscriptaddress,amsmath,amssymb]{revtex4}

\usepackage{graphicx}
\usepackage{dcolumn}
\usepackage{bm}
\usepackage{theorem}
\usepackage{color}

{\theorembodyfont{\normalfont}
\theoremheaderfont{\normalfont\it}

}

{\theorembodyfont{\normalfont}
\theoremheaderfont{\normalfont\it}
}

{\theorembodyfont{\normalfont}
\theoremheaderfont{\normalfont\it}
}

{\theorembodyfont{\normalfont}
\theoremheaderfont{\normalfont\it}
}

{\theorembodyfont{\normalfont}
\theoremheaderfont{\normalfont\it}
}

{\theorembodyfont{\normalfont}
\theoremheaderfont{\normalfont\it}
}

\newcommand{\bra}[1]{\mbox{$\left\langle#1\right|$}}
\newcommand{\ket}[1]{\mbox{$\left|#1\right\rangle$}}

\newcommand{\outpro}[2]{\mbox{$\ket{#1}\!\bra{#2}$}}
\newcommand{\proj}[1]{\mbox{$\ket{#1}\!\bra{#1}$}}

\newcommand{\prlsection}[1]{{\it{#1}}.---}

\newcommand{\beq}{\begin{eqnarray}}
\newcommand{\eeq}{\end{eqnarray}}

\begin{document}


\title{Complexity of causal order structure in distributed quantum information processing and its trade-off with entanglement}

\author{Eyuri Wakakuwa}
\affiliation{Department of Communication Engineering and Informatics, Graduate School of Informatics and Engineering, The University of Electro-Communications, Japan}%

\author{Akihito Soeda}
\affiliation{%
Department of Physics, Graduate School of Science, The University of Tokyo, Japan
}%

\author{Mio Murao}%
\affiliation{%
Department of Physics, Graduate School of Science, The University of Tokyo, Japan
}%
  \affiliation{Institute for Nano Quantum Information Electronics, The University of Tokyo, Japan}

\date{\today}

\begin{abstract}

We prove a trade-off relation between the entanglement cost and classical communication complexity of causal order structure of a protocol in distributed quantum information processing.
We consider an implementation of a class of two-qubit unitary gates by local operations and classical communication (LOCC) assisted with shared entanglement, in an information theoretical scenario  of asymptotically many input pairs and vanishingly small error. We prove the trade-off relation by showing that (i) one ebit of entanglement per pair is necessary for implementing the unitary by any two-round protocol, and that (ii) the entanglement cost by a three-round protocol is strictly smaller than one ebit per pair. We also provide an example of bipartite unitary gates for which there is no such trade-off.
\end{abstract}

\maketitle

\prlsection{Introduction}
Quantum information processing achieves its power by composing multiple quantum systems to form a larger quantum system.  It is necessary that the components collaborate to behave as a single composite quantum system.  In distributed quantum information processing (DQIP), communication channels connecting the components, quantum and/or classical, serve as a resource.  Additional correlation shared between the components is another type of resource in DQIP.  Shared correlations can be both quantum and classical.  The processing power of the individual components and the available communication/correlation resources determine the total information processing capacity of a DQIP system.

\begin{figure}[t]
\begin{center}
\includegraphics[bb={10 40 587 148}, scale=0.42]{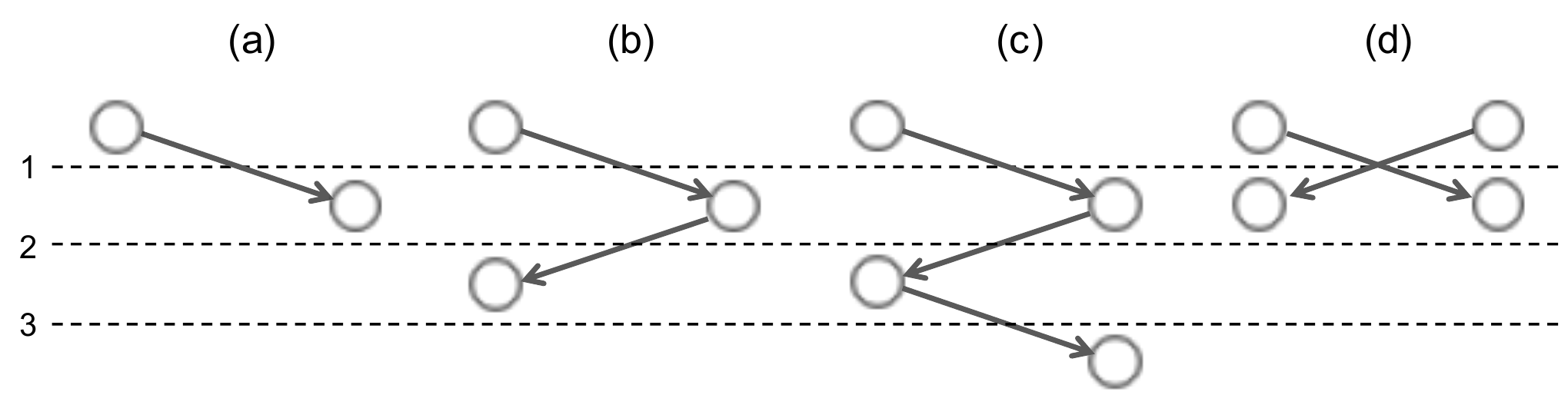}
\end{center}
\caption{Schematic description of round complexity. For each of (a)$\sim$(d), the horizontal axis represents a configuration of Alice (left) and Bob (right), and the vertical axis for time. Circles and arrows represent local operations and communications, respectively. The number of communication rounds is 1 for (a), 2 for (b), 3 for (c) and 1 for (d). (a) is a protocol with unidirectional communication, while the others are with bidirectional one. (d) is a protocol with simultaneous message exchange, while the others are not.
}
\label{fig:A}
\end{figure}

Known DQIP protocols in 
communication complexity \cite{cleve1999quantum,xue2001reducing,buhrman2016quantum,brukner2002quantum,brukner2003quantum,martinez2018high,hardy2005entanglement,kamat2008improvements,brukner2004bell,brassard2003quantum,buhrman2001quantum,buhrman2010nonlocality,tavakoli2017higher,cleve1997substituting,buhrman1999multiparty,epping2013bound}, 
interactive proof systems
\cite{kobayashi2003quantum,kobayashi2002quantum,kempe2009using,cleve2008perfect,leung2008coherent},
nonlocal games
\cite{cleve2004consequences,manvcinska2014unbounded,buscemi2012all,slofstra2011lower,kempe2010unique,kempe2011parallel,pappa2015nonlocality,kempe2010no,manvcinska2015deciding,briet2013explicit,doherty2008quantum},
measurement-based quantum computation
\cite{briegel2009measurement,walther2005experimental,raussendorf2003measurement}
and quantum cryptography \cite{gisin2002quantum} 
exhibit advantages over their classical counterparts by exploiting shared entanglement, arguably the most resourceful kind of quantum correlation.  
The round complexity (see \textit{e.g.}~Ref.~\cite{polychroniadou2016communication} and the references therein) is another type of resource inherent to a protocol.  Consider a DQIP task performed by two distant parties, say Alice and Bob.  Any protocol for this task consists of concatenations of one party performing a local operation and communicating a message to the other.  The number of concatenation is called the round complexity of a protocol (see Figure \ref{fig:A}).  Every communication must wait a certain minimum amount of time to complete, hence the round complexity of a protocol draws a lower bound on the time required.

Although the resources for DQIP have been extensively investigated \cite{bennett96,daniel03,scott07,masaki08,masaki10,masaki14,masaki15,eisert00, cirac01, groisman05, chen05, ye06, berry07, zhao08, yu10, cohen10,soeda11, stahlke11,xin08,chitambar2011local,chitambar2017round}, 
less is known about the round complexity and its relation to other resources \cite{chitambar2017round,xin08,chitambar2011local}. In an LOCC (local operations and classical communication) paradigm, all the communication is restricted to classical ones, in which case, no protocol of however high round complexity can increase the amount of entanglement.  In other words, a higher round complexity in classical communication is never a substitute for entanglement, regardless of the local processing power which only affects the set of possible LO.

\begin{figure}[t]
\begin{center}
\includegraphics[bb={40 40 658 364}, scale=0.28]{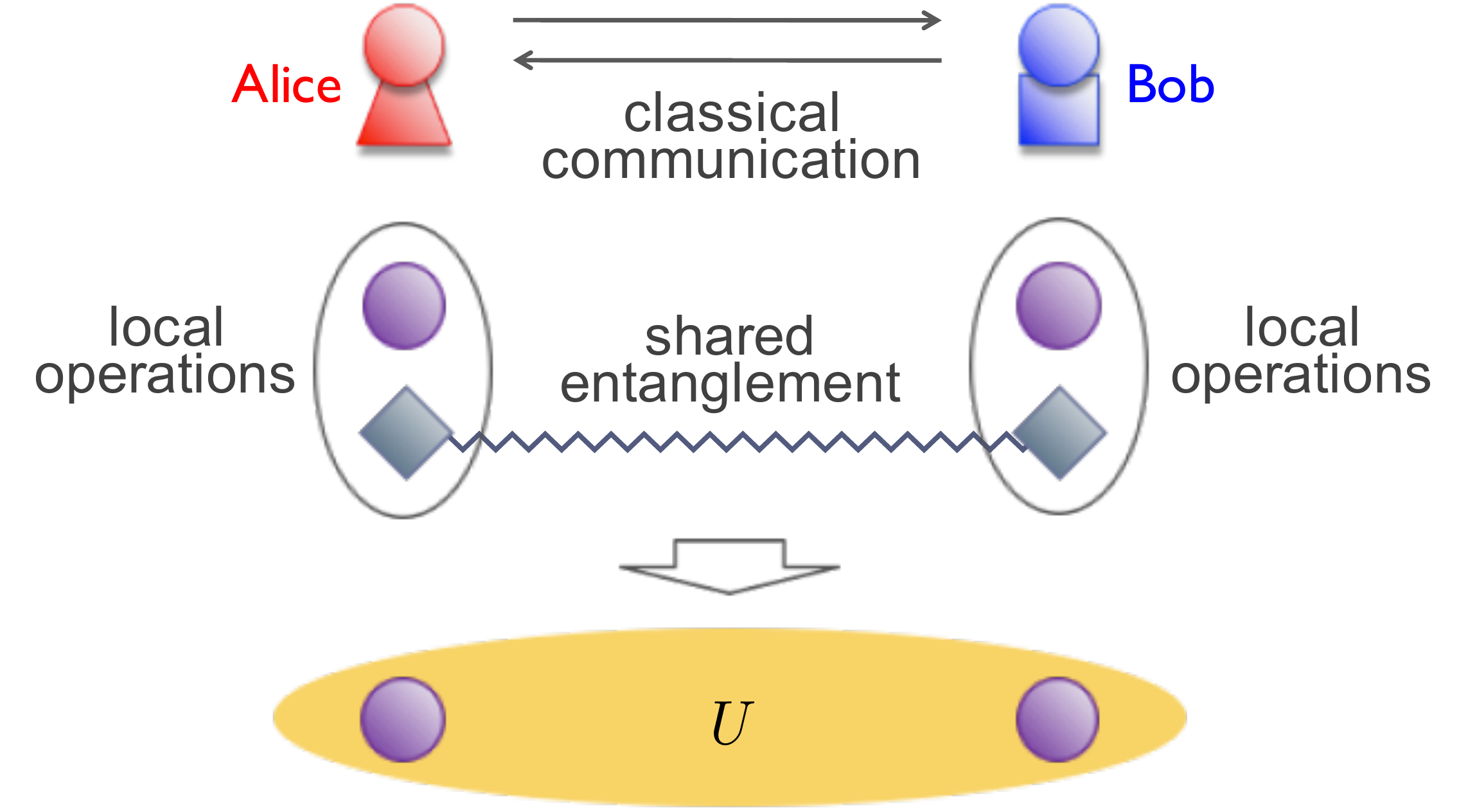}
\end{center}
\caption{Implementation of a bipartite unitary gate by LOCC assisted by shared entanglement is depicted. The balls represent physical systems on which the unitary gate is to be implemented, and the diamonds represent parts of the entanglement resource shared in advance.
}
\label{fig:B}
\end{figure}

Nevertheless, in this paper, we report a DQIP task for which the cost of shared entanglement can be reduced by increasing the CC round complexity of protocols. 
Thus, we jointly analyze entanglement and causal relations, each a fundamental topic of physics in general, in this single context of quantum information processing.
The task is for the two distant parties to implement a class of two-qubit unitary gates by LOCC assisted by shared entanglement (see Figure \ref{fig:B}).  The two parties are not allowed to exchange messages simultaneously.  We prove that a three-round protocol outperforms all two-round protocols
in reducing the entanglement cost.  Thereby we show a clear trade-off relation between the cost of entanglement and complexity in causal order.  We also provide a class of bipartite unitary gates for which there is no such trade-off, by proving that a protocol of Type (b) and (d) in Figure \ref{fig:A} achieve the minimum cost of entanglement over all finite-round protocols.

In contrast to the previous approaches \cite{bennett96,daniel03,scott07,masaki08,masaki10,masaki14,masaki15} showing advantages of bidirectional communication over unidirectional one, our result is a more ``refined'' trade-off relation between the entanglement cost and round complexity.  
Other known results \cite{eisert00, cirac01, groisman05, chen05, ye06, berry07, zhao08, yu10, cohen10,soeda11, stahlke11,xin08,chitambar2011local} analyze single-shot regimes, while we adopt an information theoretic scenario of infinitely many inputs and vanishingly small error \cite{wakakuwa2017coding}.  The more refined analysis is made possible partly due to the mathematically well-structured tools developed in quantum Shannon theory \cite{horo07,berry05,wildetext} (see \cite{wakakuwa2017coding} for the details).

\prlsection{Setup of our protocol}
Suppose that Alice and Bob, located in two distant laboratories, have $n$-qubit systems $A^n=A_1\cdots A_n$ and $B^n=B_1\cdots B_n$, respectively. They aim to apply a two-qubit unitary gate $U$ on each pair $A_iB_i\:(i=1,\cdots,n)$, simultaneously. To accomplish this task, Alice and Bob may perform quantum operations locally in their laboratories, communicate classical messages to each other, and may use copies of a Bell pair $|\Phi_2\rangle:=(|00\rangle+|11\rangle)/\sqrt{2}$ shared in advance as a resource. They are, however, not allowed to communicate quantum messages or to perform operations that globally act across their laboratories. 
That is, they accomplish the task by LOCC assisted by entanglement. We assume that they are not allowed to communicate classical messages simultaneously in both direction. The state on system $A^nB^n$ may initially be correlated with an external reference system $R$, which is inaccessible to Alice and Bob.

Let $E>0$ be the number of copies of Bell pairs divided by $n$. Denoting by $a$ and $b$ the quantum registers in which the resource state $\Phi_2$ is stored, an LOCC protocol for the above task is represented by a CPTP (completely-positive and trace-preserving) map ${\mathcal M}_n$ from $A^nB^na^{nE}b^{nE}$ to $A^nB^n$. 
The error of the protocol for a particular initial state $|\psi\rangle^{A^nB^nR}$ is quantified by the fidelity between the target state $U^{\otimes n}|\psi\rangle^{A^nB^nR}$ and the state obtained after the protocol, i.e.,
\begin{align}
\!\!\epsilon({\mathcal M}_n,\psi):=1-F\left(U^{\otimes n}(\psi)U^{\dagger \otimes n},{\mathcal M}_n(\psi\otimes\Phi_2^{\otimes nE})\right).\nonumber
\end{align} 
We adopted notations $\psi=|\psi\rangle\!\langle\psi|$ and $\Phi_2=|\Phi_2\rangle\!\langle\Phi_2|$. The fidelity is defined by $F(\rho,\sigma):=({\rm Tr}[\sqrt{\sqrt{\rho}\sigma\sqrt{\rho}}])^2$. The supremum of the above quantity over all $\psi$ is called the {\it worst-case} error and is denoted by $\epsilon^*({\mathcal M}_n)$.

An entanglement consumption rate $E$ is said to be achievable by $r$-round protocols if there exists a sequence $\{{\mathcal M}_n\}_{n=1}^\infty$ of $r$-round protocols such that the worst-case error $\epsilon^*({\mathcal M}_n)$ vanishes in the limit of $n$ to infinity. We require that the convergence of the error is sufficiently fast so that
\begin{align}
\lim_{n\rightarrow\infty}n^4\cdot\epsilon^*({\mathcal M}_n) =0.\label{eq:solace}
\end{align}
The {\it entanglement cost} of a two-qubit unitary gate $U$ by $r$-round protocols is the minimum rate $E$ that is achievable by $r$-round protocols, and is denoted by $E_r(U)$.

In this Letter, we prove that there exists a trade-off relation between the entanglement cost and round complexity for implementing a two-qubit unitary gate. 
By ``trade-off relation'', we refer to the fact that the entanglement cost of a unitary gate by the best possible $r$-round protocol is strictly smaller than the $r'$-round one, i.e., $E_r(U)<E_{r'}(U)$, for certain $r>r'$.

We consider a class of two-qubit unitary gates of the form
\begin{align}
U_\theta^{AB}=\cos{\left(\frac{\theta}{2}\right)}\cdot I^A\otimes I^B+i\sin{\left(\frac{\theta}{2}\right)}\cdot \sigma_z^A\otimes \sigma_z^B,\label{eq:atarimae}
\end{align}
where $\theta\in(0,\pi/2]$ and $I$ and $\sigma_z$ are the identity operator and the Pauli-$z$ operator defined by $I=|0\rangle\!\langle0|+|1\rangle\!\langle1|$ and $\sigma_z=|0\rangle\!\langle0|-|1\rangle\!\langle1|$, respectively. We prove that the trade-off relation holds for $\theta$ smaller than a constant, by showing that $E_2(U_\theta)>E_3(U_\theta)$.  
In the following, we describe an outline of the proof of $E_2(U_\theta)\geq1$ based on our previous work  \cite{wakakuwa2017coding}, and that of a proof of $E_3(U_\theta)<1$. A detailed proof of $E_3(U_\theta)<1$ is provided in \cite{SUPMAT}.

\prlsection{Conditions for successful protocols}
For a protocol $\mathcal{M}_n$ to be successful, the following conditions must be satisfied. 
We first analyze a general case where $A$ and $B$ are quantum systems with an arbitrary (but finite) dimension $d$. We consider a particular initial state $|\Psi_{U^\dagger\!,n}\rangle:=|\Psi_{U^\dagger}\rangle^{\otimes n}$, where $|\Psi_{U^\dagger}\rangle$ is the Choi-Jamio\l kowski state corresponding to the inverse of the unitary gate to be implemented. With $R_A$ and $R_B$ denoting $d$-dimensional reference systems that are inaccessible to Alice and Bob, the Choi-Jamio\l kowski state is defined as
\begin{align}
|\Psi_{U^\dagger}\rangle:=U^{\dagger AB}|\Phi_d\rangle^{AR_A}|\Phi_d\rangle^{BR_B},\nonumber
\end{align}
where $\Phi_d$ is the maximally entangled state with Schmidt rank $d$. 
Noting that $UU^\dagger =I$, a successful protocol ${\mathcal M}_n$ must satisfy 
\begin{align}
\!\!\!{\mathcal M}_n(\Psi_{U^\dagger\!,n}^{A^nB^nR_A^nR_B^n}\otimes\Phi_2^{\otimes nE})\approx(|\Phi_d\rangle^{AR_A}|\Phi_d\rangle^{BR_B})^{\otimes n}.\label{eq:appdec}
\end{align}
This condition imposes a restriction on Alice's measurement at the beginning the protocol. The entanglement consumption rate $E$ in a two-round protocol must be large enough in order that such a measurement by Alice exists for sufficiently large $n$.

\begin{figure}[t]
\begin{center}
\includegraphics[bb={0 30 496 196}, scale=0.42]{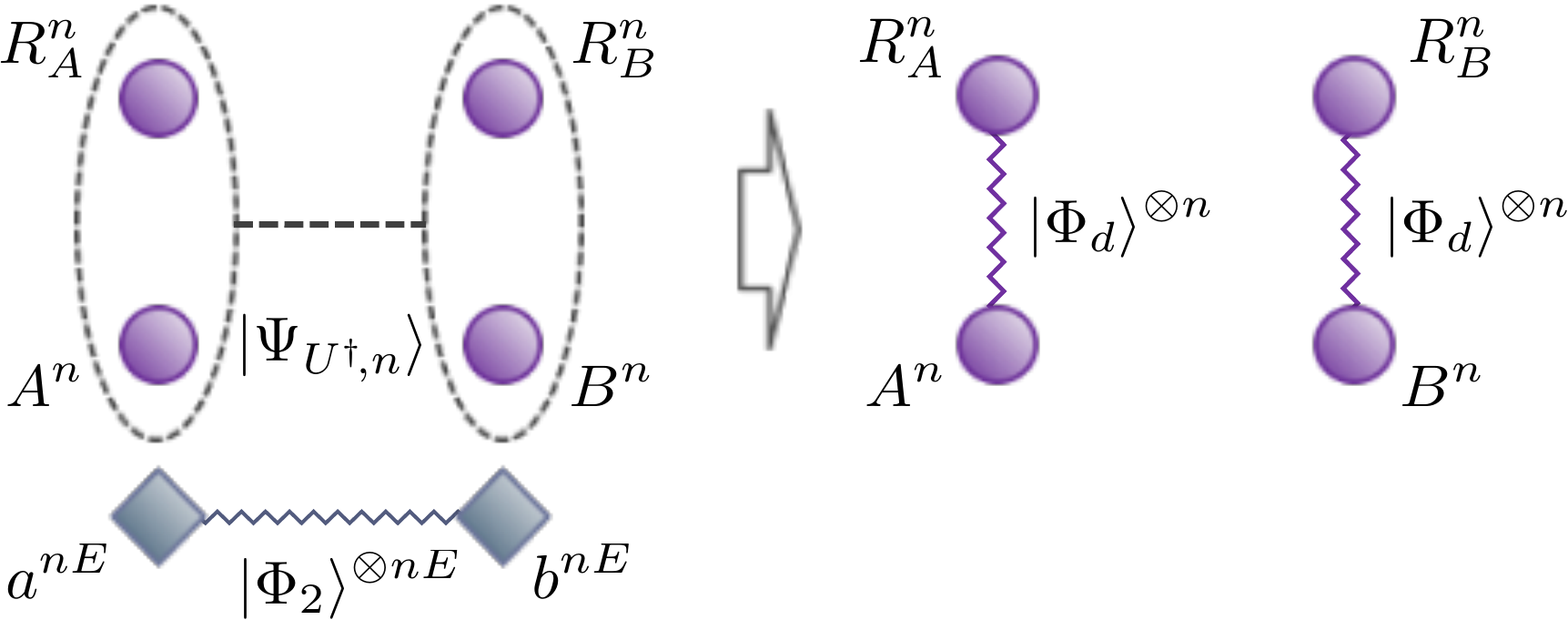}
\end{center}
\caption{A graphical representation of a task corresponding to (\ref{eq:appdec}). The task is to destroy correlation between $A^nR_A^n$ and $B^nR_B^n$, while preserving the maximal entanglement between $A^nB^n$ and $R_A^nR_B^n$ as well as the purity of the whole state.  
}
\label{fig:C}
\end{figure}

Observe that the initial state in the L.H.S. in (\ref{eq:appdec}) is an entangled state between $A^nR_A^na^{nE}/B^nR_B^nb^{nE}$, while the state in the R.H.S. is a product state in that separation. In addition, both states are pure maximally entangled states between $A^nB^n/R_A^nR_B^n$. Thus ${\mathcal M}_n$ can be viewed as a protocol that destroys correlation between $A^nR_A^na^{nE}/B^nR_B^nb^{nE}$ in the state $\Psi_{U^\dagger\!,n}^{A^nB^nR_A^nR_B^n}\otimes\Phi_2^{\otimes nE}$, while maintaining the purity of the whole state as well as the maximal entanglement between $A^nB^n/R_A^nR_B^n$ (Figure \ref{fig:C}). It should be noted that $R_A$ and $R_B$ are reference systems that are inaccessible to Alice and Bob. Thus the task considered here is different from transformation of bipartite pure states \cite{lo2001concentrating}.

Let us analyze conditions imposed by (\ref{eq:appdec}) on Alice's measurement at the beginning of a two-round protocol ${\mathcal M}_n$. We denote the output system of the measurement by $A'$. First, since entanglement between $A^nB^na^{nE}b^{nE}/R_A^nR_B^n$ are non-increasing under any step in ${\mathcal M}_n$, the reduced state on $R_A^nR_B^n$ must be close to the maximally mixed state for each measurement outcome. We call this condition as {\it obliviousness}, because it is equivalent to the condition that the measurement does not extract any information about the initial state. Second, since the reduced state on $B^nR_B^n$ is not changed by Alice's operation at the end, the maximally entangled state $(\Phi_2^{BR_B})^{\otimes n}$ must be obtained immediately after Bob's measurement. This implies that $A'R_A^n$ and $R_B^n$ must be in a product state after the measurement by Alice. We refer to this condition as {\it decoupling}. 

\prlsection{From decoupling to Markovianization}
Let $\Psi_k$ be the state after Alice's measurement corresponding to the outcome $k$. The decoupling condition is represented by the quantum mutual information as $I(A'R_A^n\!:\!R_B^n)_{\Psi_k}\approx0$ for all $k$, where $I(P:Q)_\rho:=S(\rho^P)+S(\rho^Q)-S(\rho^{PQ})$ and $S$ is the von Neumann entropy $S(\rho)=-{\rm Tr}[\rho\log{\rho}]$. 
Since $|\Phi_d\rangle^{AR_A}|\Phi_d\rangle^{BR_B}$ is the maximally entangled state between $AB/R_AR_B$, there exists a unitary $\hat U$ on $R_AR_B$ satisfying ${\hat U}^{R_AR_B}|\Phi_d\rangle^{AR_A}|\Phi_d\rangle^{BR_B}=U^{AB}|\Phi_d\rangle^{AR_A}|\Phi_d\rangle^{BR_B}$. It follows that the conditional quantum mutual information is equal to zero, i.e., $I(A'\!:\!B^n|R_A^nR_B^n)_{\Psi_k}=0$, where $I(P:Q|R)_\sigma:=S(\sigma^{PR})+S(\sigma^{QR})-S(\sigma^{R})-S(\sigma^{PQR})$. The chain rule of the quantum mutual information yields $I(A'R_A^n\!:\!R_B^n)\geq I(A'\!:\!B^nR_B^n|R_A^n)$, and consequently, we arrive at
\begin{align}
I(A'\!:\!B^nR_B^n|R_A^n)_{\Psi_k}\approx0\quad(\forall k),\label{eq:markovianize}
\end{align} 
since the conditional quantum mutual information is always non-negative \cite{lieb1973proof}.

 A tripartite quantum state for which the conditional quantum mutual information is approximately equal to zero, like (\ref{eq:markovianize}), is called an {\it approximate quantum Markov chain} (AQMC) \cite{fawzi2015quantum}.  From Condition (\ref{eq:markovianize}), it follows that Alice's measurement needs to transform the state $|\Psi_{U^\dagger\!,n}\rangle$ to an AQMC with the assistance of $|\Phi_2\rangle^{\otimes nE}$, while respecting the obliviousness condition. The entanglement consumption rate $E$ must be large enough in order that a measurement by Alice satisfying this condition exists.

\prlsection{Markovianizing cost}
We have proved in  \cite{wakakuwa2017markovianizing,wakakuwa2017cost}  that the entanglement consumption rate $E$ must be no smaller than the {\it Markovianizing cost} of $|\Psi_{U^\dagger}\rangle$ in order that there exists a measurement satisfying the condition mentioned above. In general, the Markovianizing cost of a tripartite quantum state $\rho^{ABC}$ is defined as the minimum cost of randomness required for transforming copies of the state to an approximate Markov chain, by a random unitary operation on system $A$. 
In the case of pure states, a single-letter formula for the Markovianizing cost is obtained in terms of the Koashi-Imoto decomposition \cite{koashi02}, which is used to characterize the structure of quantum Markov chains \cite{hayden04}. In the current context, the relevant Markovianizing cost is that of a `tripartite' pure state $|\Psi_{U^\dagger}\rangle$, with systems $B$ and $R_B$ treated as a single system $BR_B$.

\prlsection{Outline the proof of $E_2(U_\theta)\geq1$} 
As proven in  \cite{wakakuwa2017markovianizing,wakakuwa2017coding}, the Markovianizing cost of $\Psi_{U^\dagger}$ is equal to the von Neumann entropy of a state $\Phi_{U,\infty}^{AR_A}:=\lim_{N\rightarrow\infty}N^{-1}\sum_{n=1}^N{\mathcal E}_U^n(|\Phi_d\rangle\!\langle\Phi_d|^{AR_A})$, where ${\mathcal E}_U$ is a CPTP map on system $A$ defined by ${\mathcal E}_U(\tau)={\rm Tr}_{BR_B}[U^{AB}({\rm Tr}_B[U^{\dagger AB}(\tau^A\otimes I^B)U^{AB}]\otimes\Phi_d^{BR_B})U^{\dagger AB}]$.  For $U_\theta$ defined by (\ref{eq:atarimae}), we have ${\mathcal E}_{U_\theta}(\tau)=\frac{1}{2}((1+\cos^2{\theta})\cdot\tau+\sin^2{\theta}\cdot \sigma_z\tau \sigma_z)$ and $\Phi_{U_\theta,\infty}^{AR_A}=\frac{1}{2}(|0\rangle\!\langle0|\otimes|0\rangle\!\langle0|+|1\rangle\!\langle1|\otimes|1\rangle\!\langle1|)$. Hence the Markovianizing cost of $\Psi_{U_\theta^\dagger}$ is equal to $1$, which completes the proof of $E_2(U_\theta)\geq1$.

\prlsection{Outline of Proof of $E_3(U_\theta)<1$}
To prove $E_3(U_\theta)<1$, we first analyze a single-shot protocol proposed in  \cite{ye06} for implementing $U_\theta$. We will later extend this protocol to the one for implementing $U_\theta^{\otimes n}$, and analyze the total error and the entanglement cost by applying the law of large numbers.

The single-shot protocol consists of a concatenation of two two-round protocols and proceeds as follows: (P1)  Alice and Bob implement $U_\theta$ by a protocol of Type (b) in Figure \ref{fig:A}, using a two qubit state $|\phi_\theta\rangle^{ab}$ as a shared resource. The protocol succeeds in implementing $U_\theta$ with a certain probability $p_\theta$. If it fails, another unitary  gate $U_{\theta'}$ is implemented, in which case Alice and Bob continue to the next step. (P2) Alice and Bob implement $U_{\theta-\theta'}$ by a deterministic protocol proposed in  \cite{eisert00}, which consumes one Bell pair. The protocol is of Type (b), except that the roles of Alice and Bob are exchanged. Noting that $U_{\theta-\theta'}U_{\theta'}=U_\theta$, they succeed in implementing $U_\theta$ in total, regardless of the failure in (P1). The average entanglement cost of this protocol, measured by the entanglement entropy, is equal to ${\bar E}_\theta=1-p_\theta+E(\phi_\theta)$, where $E(\phi_\theta):=S(\phi_\theta^a)$ and $\phi_\theta^{a}:={\rm Tr}_{b}[|\phi_\theta\rangle\!\langle\phi_\theta|^{ab}]$. As we prove in \cite{SUPMAT}, it holds that  $\bar{E}_\theta<1$ for $\theta$ below a strictly positive constant.

\begin{figure}[t]
\begin{center}
\includegraphics[bb={10 40 631 202}, scale=0.38]{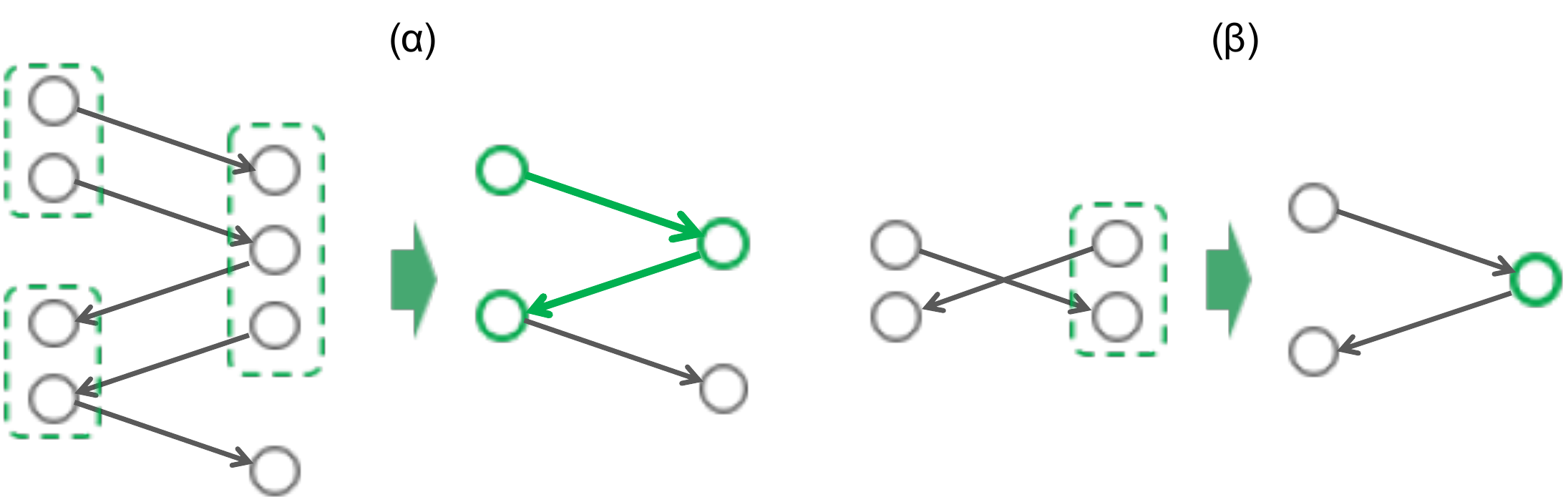}
\end{center}
\caption{Transformations of protocols in terms of communication rounds. ($\alpha$) represents how three protocols are combined to form a three-round protocol. ($\beta$) shows that a one-round protocol with simultaneous message exchange is transformed to a two-round protocol without simultaneous message exchange.}
\label{fig:E}
\end{figure}

Consider the following protocol for implementing $U_\theta^{\otimes n}$: (P0) Alice and Bob obtains $n$ copies of $|\phi_\theta\rangle^{ab}$ from approximately $nE(\phi_\theta)$ copies of Bell pairs, by an entanglement dilution protocol \cite{bennett966} of Type (a) in Figure \ref{fig:A}. (P1') They apply (P1) independently on each of $n$ input pairs. Due to the law of large numbers, $U_{\theta}$ is implemented on approximately $np_\theta$ pairs of the input. (P2') They apply (P2) to implement $U_{\theta-\theta'}$ on the remaining input pairs, which costs approximately $n(1-p_\theta)$ Bell pairs. In total, the protocol succeeds in implementing $U_\theta^{\otimes n}$ with high probability by using approximately $n{\bar E}_\theta$ copies of Bell pairs. As depicted in  Figure \ref{fig:E} ($\alpha$), the three subprotocols are brought together to form a three-round protocol. Thus it follows that $E_3(U_\theta)\leq {\bar E}_\theta$.

\prlsection{Unitaries with No Trade-off}
So far, we have investigated the case in which there is a difference between $E_2(U)$ and $E_3(U)$. Next, 
we provide an example of bipartite unitary gates for which there exists no trade-off relation between the entanglement cost and round complexity. Let $\{|t\rangle\}_{t=1}^d$ be a fixed basis of a $d$-dimensional Hilbert space $\mathcal{H}$. The generalized Pauli operators $\sigma_{pq}\:(p,q\in\{1,\cdots,d\})$ on $\mathcal{H}$ is defined by $\sigma_{pq}:=\sum_{t=1}^{d}e^{2\pi iqt/d}\outpro{t-p}{t}$, where subtraction is taken with mod $d$. Let $A$ and $B$ be $d$-dimensional systems. A bipartite unitary gate $U$ acting on $AB$ is called a generalized Clifford operator if, for any $p$, $q$, $r$ and $s$, there exist $p'$, $q'$, $r'$, $s'$ and a phase $\theta_{pqrs}\in{\mathbb R}$ such that
 \begin{eqnarray}
 U(\sigma_{pq}\otimes\sigma_{rs})U^{\dagger}=e^{i\theta_{pqrs}}\sigma_{p'q'}\otimes\sigma_{r's'}.\label{eq:makimodoshi}
 \end{eqnarray}
 In the following, we prove that $E_2(U)=\inf_{r\geq 1}E_r(U)$ holds for generalized Clifford operators. 
 
 \begin{figure}[t]
\begin{center}
\includegraphics[bb={10 40 708 218}, scale=0.30]{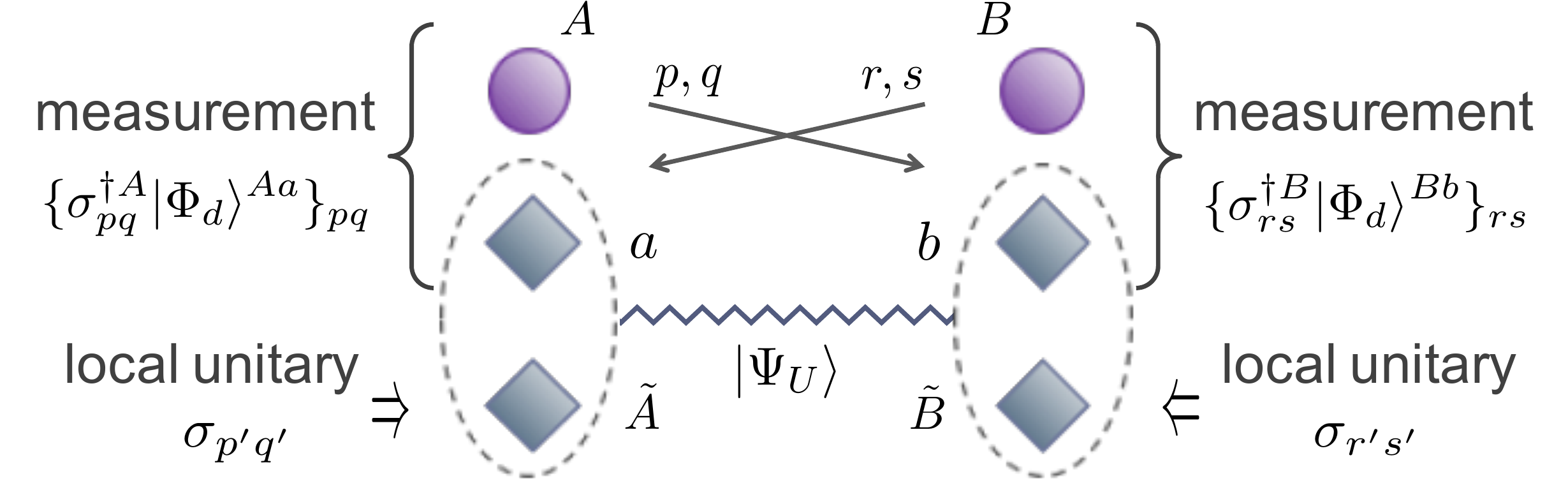}
\end{center}
\caption{A graphical representation of a single-shot protocol for implementing a generalized Clifford gate. The balls represent the input pair, and the diamonds represents parts of the shared resource. The outcomes of local measurements by Alice and Bob is simultaneously communicated to each other. $\tilde{A}$ and $\tilde{B}$ also serve as output systems of the protocol. 
}
\label{fig:D}
\end{figure}

Consider the following single-shot protocol which is depicted in Figure \ref{fig:D}: (i) Alice and Bob initially share a resource state $|\Psi_{U}\rangle^{\tilde{A}\tilde{B}ab}:=(U^{\tilde{A}\tilde{B}}\otimes I^{ab})|\Phi_d\rangle^{\tilde{A}a}|\Phi_d\rangle^{\tilde{B}b}$, with $a$ and $b$ being $d$-dimensional quantum systems; (ii) They perform a projective measurement on system $Aa$ and $Bb$ with respect to bases $\{\sigma_{pq}^{\dagger A}|\Phi_d\rangle^{Aa}\}_{pq}$ and $\{\sigma_{rs}^{\dagger B}|\Phi_d\rangle^{Bb}\}_{rs}$, respectively; (iii) They communicate the measurement outcomes $pq$ and $rs$ with each other; (iv) They perform $\sigma_{p'q'}$ on $\tilde{A}$ and $\sigma_{r's'}$ on $\tilde{B}$, respectively, determined by (\ref{eq:makimodoshi}).  This protocol is a one-round protocol of Type (d) in Figure \ref{fig:A}. A simple calculation yields that $U$ is implemented on the initial state by this protocol.

Let $K(U)$ be the entanglement entropy of $|\Psi_{U}\rangle^{\tilde{A}\tilde{B}ab}$, i.e., $K(U):=S(\Psi_{U}^{\tilde{A}a})$. Consider the following $n$-shot protocol: (i) Alice and Bob initially share approximately $nK(U)$ copies of Bell pairs, which is transformed to $n$ copies of $|\Psi_{U}\rangle^{\tilde{A}\tilde{B}ab}$ by entanglement dilution; (ii) They perform the single-shot protocol presented above on each pair. The total error of this protocol is equal to one induced in Step (i), which does not depend on the initial state and vanishes exponentially in the limit of $n$ to infinity. Since any one-round protocol with simultaneous message exchange is transformed to a two-round one without simultaneous message exchange (see Figure \ref{fig:E} ($\beta$)), it follows that $E_2(U)\leq K(U)$. The converse bound $E_r(U)\geq K(U)\;(r\geq1)$ simply follows from the monotonicity of entanglement under LOCC \cite{plenio07} for a particular initial state $(|\Phi_d\rangle^{AR_A}|\Phi_d\rangle^{BR_B})^{\otimes n}$.

Beigi et al.~\cite{beigi2011simplified} proved that any bipartite unitary gate can be implemented with arbitrary high precision by a one-round protocol of Type (d) in Figure \ref{fig:A}. The protocol proposed therein is universal in the sense that it is applicable to any type of unitary gates. The entanglement cost of the protocol, however, diverges if the total error is required to be vanishingly small. This is in contrast to the protocol presented above, which is specific to generalized Clifford gates.

\prlsection{Conclusion}
We considered implementation of a bipartite unitary gate by local operations and classical communication (LOCC), assisted by shared entanglement. We proved that a three-round protocol outperforms all two-round LOCC protocols in reducing the entanglement cost for a class of two-qubit unitary gates. Thereby we provided a first example of distributed information processing task for which there exists a clear trade-off relation between the costs of shared entanglement and the round complexity of a protocol. We also provided an example of unitary gates for which there is no such trade-off. It was proved in \cite{chitambar2011local} that a protocol with higher round complexity is more efficient in extracting entanglement from tripartite quantum state. To compare their result with a trade-off relation presented in this Letter is left as a future work.

\begin{acknowledgments}
This work is supported by the Project for Developing Innovation Systems of MEXT, Japan and JSPS KAKENHI (Grant No.~26330006, No.~15H01677, No.~16H01050, No.~17H01694, No.~18H04286,
No.~18J01329 and No.~18K13467).  We also gratefully acknowledge to the ELC project (Grant-in-Aid for Scientific Research on Innovative Areas MEXT KAKENHI (Grant No.~24106009)) for encouraging the research presented in this paper. 
\end{acknowledgments}

\newpage

\section*{SUPPLEMENTAL MATERIAL}

In this material, we provide a detailed proof of $E_3(U_\theta)<1$. We first describe a single-shot protocol for implementing $U_\theta$, which is proposed in  \cite{ye06}, and prove that the average entanglement cost is strictly smaller than $1$ for $\theta$ below a constant. Based on this protocol, we construct a protocol for implementing $U_\theta^{\otimes n}$ and evaluate the total error. A typicality argument implies that the protocol satisfies the fast convergence condition (\ref{eq:solace}).

\subsection{Single-shot protocol for $U_\theta$}\label{app:fourround}

A protocol for implementing $U_\theta$, which is proposed in  \cite{ye06}, consists of two subprotocols (P1) and (P2). In (P1), Alice and Bob uses the following state as a shared resource:
\begin{eqnarray}
|\phi_{(\alpha)}\rangle^{ab}=\cos{\left(\frac{\alpha}{2}\right)}|0\rangle|0\rangle+i\sin{\left(\frac{\alpha}{2}\right)}|1\rangle|1\rangle.\nonumber
\end{eqnarray}
Suppose that the initial state is $|\psi\rangle^{ABR}$. The protocol proceeds as follows:
\begin{enumerate}
\item Alice performs the controlled phase gate
\begin{eqnarray}
U_{\rm CZ}^{aA}=\proj{0}^{a}\otimes I^A+\proj{1}^{a}\otimes \sigma_z^A,\nonumber
\end{eqnarray}
after which the whole state is
\begin{eqnarray}
|\psi_{tot}'\rangle^{abABR}&=&\cos{\left(\frac{\alpha}{2}\right)}|0\rangle^{a}|0\rangle^{b}|\psi\rangle^{ABR}\nonumber\\
&&\!\!\!\!\!\!\!\!+i\sin{\left(\frac{\alpha}{2}\right)}|1\rangle^{a}|1\rangle^{b}\sigma_z^A|\psi\rangle^{ABR}.\nonumber
\end{eqnarray}
\item Alice performs a projective measurement on $a$ with basis $\{|+\rangle,|-\rangle\}$, and sends the outcome to Bob.
\item Bob performs $I$ or $\sigma_z$ on $b$ depending on the measurement outcome. The whole state is then
\begin{eqnarray}
|\psi_{tot}''\rangle^{bABR}&=&\cos{\left(\frac{\alpha}{2}\right)}|0\rangle^{b}|\psi\rangle^{ABR}\nonumber\\
&&+i\sin{\left(\frac{\alpha}{2}\right)}|1\rangle^{b}\sigma_z^A|\psi\rangle^{ABR}.\nonumber
\end{eqnarray}
\item Bob performs the controlled-$z$ gate
\begin{eqnarray}
U_{\rm CZ}^{bB}=\proj{0}^{b}\otimes I^B+\proj{1}^{b}\otimes \sigma_z^B,\nonumber
\end{eqnarray}
after which the whole state is
\begin{eqnarray}
|\psi_{tot}'''\rangle^{bABR}&=&\cos{\left(\frac{\alpha}{2}\right)}|0\rangle^{b}|\psi\rangle^{ABR}\nonumber\\
&&\!\!\!\!\!\!\!\!\!+i\sin{\left(\frac{\alpha}{2}\right)}|1\rangle^{b}(\sigma_z^A\otimes \sigma_z^B)|\psi\rangle^{ABR}.\nonumber
\end{eqnarray}
\item Bob performs a projective measurement on $b$ with basis $\{|\chi\rangle/\langle\chi|\chi\rangle^{1/2},|\chi^\perp\rangle/\langle\chi^\perp|\chi^\perp\rangle^{1/2}\}$, and sends the outcome to Alice. Here, $|\chi\rangle$ and $|\chi^\perp\rangle$ are supernormalized state vectors defined by
\begin{eqnarray}
&&|\chi\rangle:=\frac{\cos{(\theta/2)}}{\cos{(\alpha/2)}}|0\rangle+\frac{\sin{(\theta/2)}}{\sin{(\alpha/2)}}|1\rangle,\nonumber\\
&&|\chi^\perp\rangle:=\frac{\sin{(\theta/2)}}{\sin{(\alpha/2)}}|0\rangle-\frac{\cos{(\theta/2)}}{\cos{(\alpha/2)}}|1\rangle.\nonumber
\end{eqnarray}
\end{enumerate}
If the measurement outcome corresponding to $|\chi\rangle$ is obtained, the state becomes
\begin{eqnarray}
&&\!\!\!\!\!\!|\psi^s\rangle^{ABR}=\langle\chi|\psi_{tot}'''\rangle\nonumber\\
&&=\cos{\left(\frac{\theta}{2}\right)}|\psi\rangle^{ABR}+i\sin{\left(\frac{\theta}{2}\right)}(\sigma_z^A\otimes \sigma_z^B)|\psi\rangle^{ABR}\nonumber
\end{eqnarray}
as desired. The success probability is given by 
\begin{eqnarray}
p(\alpha,\theta)=\frac{|\langle\chi|\psi_{tot}'''\rangle|^2}{\langle\chi|\chi\rangle}=\frac{1}{\langle\chi|\chi\rangle}=\frac{\sin^2{\alpha}}{2(1-\cos{\theta}\cos{\alpha})}.\nonumber
\end{eqnarray}
If the complementary outcome is obtained, then the state changes
\begin{align}
&|\psi^f\rangle^{ABR}=\langle\chi^\perp|\psi_{tot}'''\rangle\nonumber\\
&=\frac{\sin{(\theta/2)}}{\tan{(\alpha/2)}}|\psi\rangle^{ABR}+i\frac{\cos{(\theta/2)}}{\tan{(\alpha/2)}^{-1}}(\sigma_z^A\otimes \sigma_z^B)|\psi\rangle^{ABR},\nonumber
\end{align}
up to normalization condition. It is straightforward to verify that the normalized state satisfies
\begin{align}
\frac{|\psi^f\rangle^{ABR}}{\||\psi^f\rangle^{ABR}\|}=U_{\theta'}|\psi\rangle^{ABR}\nonumber
\end{align}
with $\theta'$ defined by
\begin{eqnarray}
\tan{\left(\frac{\theta'}{2}\right)}=\frac{\tan^2{(\alpha/2)}}{\tan{(\theta/2)}}.\nonumber
\end{eqnarray}
In the latter case, Alice and Bob continue to (P2).

In (P2), Alice and Bob implement $U_{\theta-\theta'}$ by a protocol for implementing a two-qubit controlled-unitary gate, which is proposed in  \cite{eisert00}. (For later convenience, we exchange the roles of Alice and Bob in the original formulation.) Note that $U_{\theta-\theta'}$ is equivalent to the following controlled-unitary gate up to local unitary transformations:
\begin{align}
\tilde{U}_\theta^{AB}=I^A\otimes \proj{0}^B+(e^{i(\theta-\theta')\sigma_z})^A\otimes \proj{1}^B.
\end{align}
The protocol consumes one Bell pair $|\Phi_2\rangle^{ab}$ and deterministically implements $U_{\theta-\theta'}$. Suppose that the initial state is $|\tilde{\psi}\rangle^{ABR}$ and is decomposed in the form of
\begin{align}
|\tilde{\psi}\rangle=c_0\ket{0}^B\ket{\psi_0}^{AR}+c_1\ket{1}^B\ket{\psi_1}^{AR},
\end{align}
where $\ket{\psi_0},\ket{\psi_1}$ are normalized pure states and $c_0,c_1$ are complex numbers that satisfy $|c_0|^2+|c_1|^2=1$.  The protocol proceeds as follows:
\begin{enumerate}
\item Bob performs the CNOT gate
\begin{eqnarray}
U_{\rm CN}^{bB}= I^b\otimes\proj{0}^{B}+\sigma_x^b\otimes \proj{1}^{B},\nonumber
\end{eqnarray}
after which the whole state is
\begin{align}
|\psi_{\theta'}'\rangle^{abABR}=&\frac{c_0}{\sqrt{2}}(|0\rangle^{a}|0\rangle^{b}+|1\rangle^{a}|1\rangle^{b})\ket{0}^B\ket{\psi_0}^{AR}\!\!\!\!\!\!\nonumber\\
&\!\!\!+\frac{c_1}{\sqrt{2}}(|0\rangle^{a}|1\rangle^{b}+|1\rangle^{a}|0\rangle^{b})\ket{1}^B\ket{\psi_1}^{AR}.\!\!\!\!\!\!\!\!\!\!\!\!\!\!\!\!\!\!\!\!\!\!\!\!\!\!\!\nonumber
\end{align}
\item Bob performs a projective measurement on $b$ with basis $\{|0\rangle,|1\rangle\}$, and sends the outcome to Alice.
\item Alice performs $I$ or $\sigma_x$ on $a$ depending on the measurement outcome. The whole state is then
\begin{align}
|\psi_{\theta'}''\rangle^{abABR}=c_0|0\rangle^{a}\ket{0}^B\ket{\psi_0}^{AR}+c_1|1\rangle^{a}\ket{1}^B\ket{\psi_1}^{AR}.\!\!\!\!\!\!\!\!\!\!\!\!\!\!\!\nonumber
\end{align}
\item Alice performs the controlled-$z$ gate
\begin{align}
U^{aA}=\proj{0}^{a}\otimes I^A+\proj{1}^{a}\otimes(e^{i(\theta-\theta')\sigma_z})^A,\nonumber
\end{align}
after which the whole state is
\begin{align}
|\psi_{\theta'}'''\rangle^{bAB}=&\:c_0|0\rangle^{a}\ket{0}^B\ket{\psi_0}^{AR}\nonumber\\
&+c_1|1\rangle^{a}\ket{1}^B(e^{i(\theta-\theta')\sigma_z})^A\ket{\psi_1}^{AR}.\nonumber
\end{align}
\item Alice performs a projective measurement on $a$ with basis $\{|+\rangle,|-\rangle\}$ and sends the outcome to Bob, where $|\pm\rangle:=(\ket{0}\pm\ket{1})/\sqrt{2}$. 
\item Bob performs $I$ or $\sigma_z$ on $B$ depending on the outcome, after which the state is $\tilde{U}_{\theta-\theta'}|\tilde{\psi}\rangle^{ABR}$ as desired.
\end{enumerate}

 In total, the entanglement cost of the composite protocol of (P1) and (P2) is given by
\begin{align}
\bar{E}(\alpha,\theta)&=(1-p(\alpha,\theta))E(\Phi_2)+E(\phi_{(\alpha)})\nonumber\\
&=1-p(\alpha,\theta)+E(\phi_{(\alpha)})\nonumber
\end{align}
on average. Here, $E$ denotes the entanglement entropy defined by $E(\phi_{(\alpha)}):=S(\phi_{(\alpha)}^a)$ and $E(\Phi_2):=S(\Phi_2^a)=1$.

For a particular choice $\alpha=\sqrt{\theta}$, the resource state and the success probability in (P1) are given by
\begin{align}
|\phi_{\theta}\rangle^{ab}:=|\phi_{(\sqrt{\theta})}\rangle^{ab}&=\cos{\left(\!\frac{\sqrt{\theta}}{2}\!\right)}|0\rangle|0\rangle+i\sin{\left(\!\frac{\sqrt{\theta}}{2}\!\right)}|1\rangle|1\rangle,\nonumber\\
p_\theta:=p(\sqrt{\theta},\theta)&=\frac{\sin^2{\sqrt{\theta}}}{2(1-\cos{\theta}\cos{\sqrt{\theta}})},\nonumber
\end{align}
respectively, and the average entanglement cost is
\begin{align}
\bar{E}_\theta:=\bar{E}(\sqrt{\theta},\theta)=1-p_\theta+h(\cos^2(\sqrt{\theta}/2)),\label{eq:ethetafor}
\end{align}
where $h$ is the binary entropy defined by $h(x):=-x\log{x}-(1-x)\log{(1-x)}$. It is straightforward to verify that
\begin{align}
\lim_{\theta\rightarrow0}h(\cos^2(\sqrt{\theta}/2))=0.\label{eq:hoshizora}
\end{align}
For $\theta\approx0$, we have
\begin{eqnarray}
&&\sin^2{\sqrt{\theta}}=\theta+O(\theta^2),\nonumber\\
&&\cos{\theta}=1-\frac{1}{2}\theta^2+O(\theta^4),\nonumber\\
&&\cos{\sqrt{\theta}}=1-\frac{1}{2}\theta+O(\theta^2),\nonumber\\
&&\cos{\theta}\cos{\sqrt{\theta}}=1-\frac{1}{2}\theta+O(\theta^2).\nonumber
\end{eqnarray}
Thus we have
\begin{align}
p_\theta=\frac{\theta+O(\theta^2)}{2\left(\frac{1}{2}\theta+O(\theta^2)\right)}=1+O(\theta),\nonumber
\end{align}
which leads to
\begin{align}
\lim_{\theta\rightarrow0}p_\theta=1.\label{eq:pthetaconv}
\end{align}
From (\ref{eq:ethetafor}), (\ref{eq:hoshizora}) and (\ref{eq:pthetaconv}), we obtain $\lim_{\theta\rightarrow0}\bar{E}_\theta=0$. Since $\bar{E}_\theta$ is a continuous function of $\theta$, it follows that $\bar{E}_\theta<1$ for $\theta$ below a certain   strictly positive constant.

\subsection{Protocol for implementing $U_\theta^{\otimes n}$}

Let us turn to a protocol for implementing $U_\theta^{\otimes n}$. Fix arbitrary $\delta>0$ and choose sufficiently large $n\in{\mathbb N}$. The protocol proceeds as follows:
\begin{enumerate}
\item Alice and Bob initially share $n(\bar{E}_\theta+2\delta)=n(1-p_\theta+h_\theta+2\delta)$ copies of Bell pairs, where we denoted $h(\cos^2(\sqrt{\theta}/2))$ simply by $h_\theta$.
\item By entanglement dilution  \cite{bennett966}, they transform $n(h_\theta+\delta)$ copies of Bell pairs to a state $|\omega_n\rangle$, which is equal to $|\phi_\theta\rangle^{\otimes n}$ up to a small error $\epsilon_n:=\||\omega_n\rangle\!\langle\omega_n|-|\phi_\theta\rangle\!\langle\phi_\theta|^{\otimes n}\|_1$. The dilution protocol is a one-round protocol of Type (a) in Figure \ref{fig:A} in the main text.
\item By using $n$ copies of $|\phi_\theta\rangle$ as resources, they perform $U_\theta$ on each  of the input sequence by (P1). Either of the following two events will occur:
\begin{enumerate}
\item With a high probability, the number of pairs for which $U_\theta$ has been applied is not smaller than $n(p_\theta-\delta)$. $U_{\theta'}$ has been applied on the other pairs, the number of which is not greater than $n(1-p_\theta+\delta)$.  
\item With small probability $\epsilon_n'$, the number of pairs for which $U_\theta$ has been applied is smaller than $n(p_\theta-\delta)$.
\end{enumerate}
Continue to the next step if (a) has occurred.
\item By using the remaining $n(1-p_\theta+\delta)$ Bell pairs, they perform $U_{\theta-\theta'}$ by (P2) on pairs for which $U_{\theta'}$ has been applied.
\end{enumerate}
Note that the second-round communication in Step 3 and the first-round one in Step 4 can be performed simultaneously. 
Similarly,   the communication from Alice to Bob in Step 3 can be performed simultaneously with the communication in Step 2 as well. Hence the above protocol is organized into a three-round protocol of Type (c) in Figure \ref{fig:A}.

Let ${\mathcal M}_n'$ be a quantum operation that represents Step 3 and 4 in the above protocol, and suppose the initial state is $|\Psi_n\rangle^{A^nB^nR^n}$. The total error is evaluated as follows. If (a) occurs in Step 3, the final state is exactly equal to the target state $|\Psi_{n,{\rm tar}}\rangle:=U_\theta^{\otimes n}|\Psi_n\rangle$. Let $\tau_{(b)}$ be the state obtained when (b) occurs. The final state is, in total, given by
\begin{align}
{\mathcal M}_n'\left(\Psi_n\!\otimes\!\phi_{\theta}^{\otimes n}\!\otimes\!\Phi_2^{\otimes n(1-p_\theta+\delta)}\right)=(1-\epsilon_n')\Psi_{n,{\rm tar}}+\epsilon_n'\tau_{(b)},\nonumber
\end{align}
which leads to
\begin{align}
&\left\|{\mathcal M}_n'\left(\Psi_n\otimes\phi_\theta^{\otimes n}\otimes\Phi_2^{\otimes n(1-p_\theta+\delta)}\right)-U_\theta^{\otimes n}\Psi_nU_\theta^{\dagger \otimes n}\right\|_1\nonumber\\
&=\epsilon_n'\left\|\Psi_{n,{\rm tar}}-\tau_{(b)}\right\|_1\leq2\epsilon_n'.\label{eq:ashla}
\end{align}

Let ${\mathcal M}_n$ be a quantum operation that represents Step  2$\sim$4. By definition, we have
\begin{align}
\!\!\!{\mathcal M}_n\left(\Psi_n\otimes\Phi_{K_n}\right)={\mathcal M}_n'\left(\Psi_n\otimes\omega_n\otimes\Phi_2^{\otimes n(1-p_\theta+\delta)}\right).\!\label{eq:spp}
\end{align}
A simple calculation yields 
\begin{eqnarray}
&&\left\|{\mathcal M}_n(\Psi_n\otimes\Phi_{K_n})-U_\theta^{\otimes n}\Psi_nU_\theta^{\dagger \otimes n}\right\|_1\nonumber\\
&=&\left\|{\mathcal M}_n'\left(\Psi_n\otimes\omega_n\otimes\Phi_2^{\otimes n(1-p_\theta+\delta)}\right)-U_\theta^{\otimes n}\Psi_nU_\theta^{\dagger \otimes n}\right\|_1\nonumber\\
&\leq&\left\|{\mathcal M}_n'\left(\Psi_n\otimes\omega_n\otimes\Phi_2^{\otimes n(1-p_\theta+\delta)}\right)\right.\nonumber\\
&&\;\;\left.-{\mathcal M}_n'\left(\Psi_n\otimes\phi_\theta^{\otimes n}\otimes\Phi_2^{\otimes n(1-p_\theta+\delta)}\right)\right\|_1\nonumber\\
&&\!\!+\left\|{\mathcal M}_n'\left(\Psi_n\otimes\phi_\theta^{\otimes n}\otimes\Phi_2^{\otimes n(1-p_\theta+\delta)}\right)\right.\nonumber\\
&&-\;\;\left.U_\theta^{\otimes n}\Psi_nU_\theta^{\dagger \otimes n}\right\|_1\nonumber\\
&\leq&\left\|\Psi_n\otimes\omega_n\otimes\Phi_2^{\otimes n(1-p_\theta+\delta)}\right.\nonumber\\
&&\;\;\left.-\Psi_n\otimes\phi_\theta^{\otimes n}\otimes\Phi_2^{\otimes n(1-p_\theta+\delta)}\right\|_1+2\epsilon_n'\nonumber\\
&=&\left\|\omega_n-\phi_\theta^{\otimes n}\right\|_1+2\epsilon_n'\nonumber\\
&=&\epsilon_n+2\epsilon_n'.\nonumber
\end{eqnarray}
Here, the first equality follows from (\ref{eq:spp}); the first inequality due to the triangle inequality for the trace distance; the second inequality from the monotonicity of the trace distance and Inequality (\ref{eq:ashla}); the second equality because we have $\|\rho\otimes\tau-\sigma\otimes\tau\|_1=\|\rho-\sigma\|_1$; and the last line from the definition of $\epsilon_n$. Noting that the fidelity and the trace distance satisfy the relation $1-\sqrt{F(\rho,\sigma)}\leq\frac{1}{2}\|\rho-\sigma\|_1$ (see e.g.  \cite{wildetext}), it follows that the protocol ${\mathcal M}_n$ satisfies
\begin{align}
\epsilon^*(\mathcal{M}_n)=\sup_{\Psi_n}\:\epsilon(\mathcal{M}_n,\Psi_n)\leq\epsilon_n+2\epsilon_n'.\nonumber
\end{align} 
As we prove in the next section, $\epsilon_n$ and $\epsilon_n'$ converges to zero exponentially with $n$. Thus the protocol $\mathcal{M}_n$ above satisfies the fast convergence condition (\ref{eq:solace}).

\subsection{Evaluation of $\epsilon_n$ and $\epsilon_n'$}\label{app:prfdisti}

Define
\begin{align}
\lambda_0=\cos^2\left(\frac{\sqrt{\theta}}{2}\right),\;\lambda_1=\sin^2\left(\frac{\sqrt{\theta}}{2}\right),\nonumber
\end{align}
and fix arbitrary $\delta>0$, $n\in{\mathbb N}$. A sequence ${\bm x}=(x_1,\cdots,x_n)\in\{0,1\}^n$ is said to be {\it$\delta$-weakly typical with respect to} $\{\lambda_x\}_{x\in\{0,1\}}$ if it satisfies
\begin{eqnarray}
2^{-n(H(\{\lambda_x\})+\delta)}\leq\prod_{i=1}^n\lambda_{x_i}\leq2^{-n(H(\{\lambda_x\})-\delta)}.\label{eq:kinoshitade}
\end{eqnarray}
Here, $H(\{\lambda_x\})$ is the Shannon entropy of a probability distribution $\{\lambda_x\}_{x\in\{0,1\}}$ defined by
\begin{align}
H(\{\lambda_x\}):=-\sum_{x=\{0,1\}}\lambda_x\log{\lambda_x},\nonumber
\end{align}
and is equal to $h_\theta$. The set of all $\delta$-weakly typical sequences is called the {\it$\delta$-weakly typical set}, and is denoted by ${\mathcal T}_{n,\delta}$. The {\it$\delta$-weakly typical subspace of $({\mathcal H}^{a})^{\otimes n}$ with respect to $\phi_\theta^{a}={\rm Tr}_{b}[|\phi_\theta\rangle\!\langle\phi_\theta|^{ab}]$} is defined as
\begin{eqnarray}
&&\!\!\!\!\!\!\!\!\!\!\!{\mathcal H}_{n,\delta}:=\nonumber\\
&&\!\!\!\!\!\!{\rm span}\left\{\left.\ket{x_1}\cdots\ket{x_n}\in({\mathcal H}^{a})^{\otimes n}\right|(x_1,\cdots,x_n)\in{\mathcal T}_{n,\delta}\right\}.\!\!\!\!\!\!\!\!\!\nonumber
\end{eqnarray}
Let $\Pi_{n,\delta}$ be the projection onto ${\mathcal H}_{n,\delta}\subseteq({\mathcal H}^{a})^{\otimes n}$, and let us introduce a notation 
\begin{align}
\lambda_{\bm x}:=\lambda_{x_1}\cdots\lambda_{x_n}.\nonumber
\end{align}
 Abbreviating $(\Pi_{n,\delta}\otimes I^{b^n})|\phi_\theta\rangle^{\otimes n}$ as $\Pi_{n,\delta}|\phi_\theta\rangle^{\otimes n}$, we have
\begin{align}
{\rm Tr}[\Pi_{n,\delta}(|\phi_\theta\rangle\!\langle\phi_\theta|^{\otimes n})]=\sum_{{\bm x}\in{\mathcal T}_{n,\delta}}\lambda_{\bm x}.\label{eq:a}
\end{align}

It is proved in  \cite{ahlswede80} that there exists a constant $c>0$, which depends on ${\{\lambda_x\}_x}$, such that for any $\delta>0$ and $n$, we have
\begin{eqnarray}
\sum_{{\bm x}\in{\mathcal T}_{n,\delta}}\lambda_{\bm x}\geq1-\exp{(-c\delta^2n)}.\nonumber
\end{eqnarray}
Denoting this constant by $c_\theta$, we obtain
\begin{align}
{\rm Tr}[\Pi_{n,\delta}(|\phi_\theta\rangle\!\langle\phi_\theta|^{\otimes n})]\geq1-\exp{(-c_\theta\delta^2n)}.\label{eq:expconv}
\end{align}

Fix arbitrary $\delta>0$, $n\in{\mathbb N}$, and consider the normalized state $|\omega_n\rangle$ defined by
\begin{eqnarray}
|\omega_n\rangle:=\frac{\Pi_{n,\delta}(|\phi_\theta\rangle^{\otimes n})}{\sqrt{{\rm Tr}[\Pi_{n,\delta}(|\phi_\theta\rangle\!\langle\phi_\theta|^{\otimes n})]}}.\nonumber
\end{eqnarray}
Due to the gentle measurement lemma (see e.g. Lemma 9.4.1 in  \cite{wildetext}) and Ineq.~(\ref{eq:expconv}), the state satisfies
\begin{eqnarray}
\epsilon_n:=\left\||\omega_n\rangle\!\langle\omega_n|-|\phi_\theta\rangle\!\langle\phi_\theta|^{\otimes n}\right\|_1\leq2\exp{\left(-\frac{c_\theta\delta^2n}{2}\right)},\nonumber
\end{eqnarray}
where the trace distance is defined by $\|\rho-\sigma\|_1:={\rm Tr}[\sqrt{(\rho-\sigma)^2}]$. 
By definition, the Schmidt decomposition of $|\omega_n\rangle$ is given by
\begin{eqnarray}
|\omega_n\rangle=\sum_{{\bm x}\in{\mathcal T}_{n,\delta}}\sqrt{\lambda_{\bm x}'}|{\bm x}\rangle|{\bm x}\rangle,\nonumber
\end{eqnarray}
where
\begin{eqnarray}
\lambda_{\bm x}':=\frac{\lambda_{\bm x}}{{\rm Tr}[\Pi_{n,\delta}(|\phi_\theta\rangle\!\langle\phi_\theta|^{\otimes n})]}.\nonumber
\end{eqnarray}
From (\ref{eq:kinoshitade}), it follows that
\begin{eqnarray}
\lambda_{\bm x}'\geq2^{-n(H(\{\lambda_x\})+\delta)}.\nonumber
\end{eqnarray}
Thus a uniform distribution on a set $\{1,\cdots,2^{n(H(\{\lambda_x\})+\delta)}\}$ is majorized by a probability distribution $\{\lambda_{\bm x}'\}_{{\bm x}\in{\mathcal T}_{n,\delta}}$. Consequently, there exists an LOCC protocol that transforms $n(H(\{\lambda_x\})+\delta)$ copies of Bell pairs to $|\omega_n\rangle$ deterministically and exactly \cite{nielsen99}.

The law of large numbers implies $\lim_{n\rightarrow\infty}\epsilon_n'=0$. It is proved in  \cite{ahlswede80} that there exists an $n$-independent positive constant $c_\theta'$ such that
\begin{align}
\epsilon_n'\leq \exp{(-c_\theta'\delta^2n)}\nonumber
\end{align}
for any $\delta$ and $n$.

\bibliography{bibbib.bib}

\end{document}